\documentclass[submission, Phys]{SciPost}
\usepackage{graphicx}
\usepackage{amssymb,amsmath}
\usepackage[utf8]{inputenc}

\DeclareGraphicsExtensions{.png,.jpg,.bmp}

\begin{document}
\begin{center}{\Large \textbf{
Design Rules for DMI-Stabilised Skyrmions
}}\end{center}

\begin{center}
L. Ranno\textsuperscript{1*},
M. Alfonso Moro\textsuperscript{1}
\end{center}

\begin{center}
{\bf 1} Institut N\'eel, CNRS/Universit\'e Grenoble Alpes, Grenoble, France
\\

* laurent.ranno@neel.cnrs.fr
\end{center}

\begin{center}
\today
\end{center}


\section*{Abstract}
{\bf
Magnetic skyrmions are topological objects, which have recently been observed in thin films at room temperature. Sub 100-nm sizes and spin polarised current manipulation make them candidates for high density information storage and processing. Besides material thickness, skyrmion stability involves many contributions: ferromagnetic exchange, magnetic anisotropy, Dzyaloshinskii-Moryia exchange interaction (DMI) and demagnetising energy. Multidimensional phase diagrams have been reported in the litterature. In this work we propose a simple two-parameter model, exact for thin films in the negligible magnetisation limit, e.g. ferrimagnets close to compensation and synthetic artificial antiferromagnets. It allows to define design rules to stabilize small diameter skyrmions at room temperature with explicit estimates of their nucleation and collapse energies. Comparisons to micromagnetic calculations allow to assess the model quality for non-zero magnetisation and show good agreement.  
}

\vspace{10pt}
\noindent\rule{\textwidth}{1pt}
\tableofcontents\thispagestyle{fancy}
\noindent\rule{\textwidth}{1pt}
\vspace{10pt}

\section{Introduction}
\label{sec:intro}

Following the $\mu$m-size magnetic bubble studies from the 1970s \cite{chaudhari_amorphous_1973, bobeck_magnetic_1975, eschenfelder_magnetic_1981}, magnetic skyrmions with a similar topology but smaller sizes have been experimentally evidenced, at first at low temperature and in bulk samples \cite{muhlbauer_skyrmion_2009,yu_real-space_2010} and then in metallic ultrathin films at room temperature \cite{jiang_blowing_2015, boulle_room-temperature_2016}. Magnetic bubbles are stabilised by their demagnetising energy whereas the skyrmion stabilisation energy is based on an asymmetric exchange interaction such as the Dzyaloshinskii Moriya Interaction (DMI) which favors whirling spin textures. This interaction also selects a whirling chirality and makes skyrmions homochiral objects \cite{bogdanov_thermodynamically_1994}. Due to their small size and the metallic character of the materials, skyrmions are being widely studied as candidates for high density information storage associated with current driven manipulation \cite{sampaio_nucleation_2013}.  

	The main characteristics of $\mu$m-size bubbles (radius, energy) can be evaluated using a few reduced quantities. For example, their characteristic lengthscale is the material length $L=\frac{4\sqrt{AK}}{\mu_0M_s^2}$, where $L$ is the ratio between the Bloch domain wall surface energy and the magnetostatic characteristic energy. The most stable bubbles \cite{debonte_properties_1973} have then a diameter 8$L$, in a film with preferred thickness t=4$L$  i.e., an aspect ratio 2. Diameters in the sub-micron range were achieved \cite{chaudhari_amorphous_1973} but further downscaling was not expected and semiconducting memories rapidly outperformed bubble memories.

	Several skyrmion-bubble phase diagrammes have recently been reported\cite{kiselev_chiral_2011,buttner_theory_2018,bernand-mantel_skyrmion-bubble_2018}. However some restrictions are usually present, for example modelling only Bloch skyrmions\cite{kiselev_chiral_2011} or Néel skyrmions with imposed profiles such as double walls\cite{buttner_theory_2018} or zero-thickness walls \cite{bernand-mantel_skyrmion-bubble_2018}, non physical zero-radius limit \cite{buttner_theory_2018}. When long-range demagnetising effects are present, these diagrammes evidence the coexistence of magnetostatically-stabilised skyrmionic "large" bubbles and DMI-stabilised "small" skyrmions, with a possible continuous path to transform one into the other \cite{bernand-mantel_skyrmion-bubble_2018}. These results lack simplicity due to the multi dimensional phase space and require heavy demagnetising field computing. Here we propose a simple model with good physical insight and apply it to define design rules to guide the fabrication of the relevant multilayers where thickness t, exchange constant A, DMI constant D, magnetisation M$_s$ and uniaxial anisotropy K$_u$ are the parameters to be chosen in order to stabilise skyrmions with 10 nm diameter, stable at room temperature.\\

\section{Model}

	The modeled material is an ultra-thin magnetic film, with thickness t (typically 0.5 to 5 nm, along the z-axis). It is ferromagnetic at room temperature (exchange constant A, typically 10 pJ/m), its spontaneous magnetisation is $M_s$ and its uniaxial perpendicular anisotropy is K$_u$(J/m$^3$). For a thin film the demagnetising energy is characterised by $K_d=\frac{\mu_0M_s^2}{2}$. Spontaneous perpendicular magnetisation implies to keep the quality factor $Q=\frac{K_u}{K_d}$  larger than 1. DMI energy  is characterised by the DMI constant D(J/m$^2$). D values larger than a critical  value $D_c$ favor a non ferromagnetic cycloidal ground state. 

	Demagnetising energy, when non-zero, is the most complicated energy to calculate due to its non local character. For ultrathin films, a local anisotropy assumption \cite{tarasenko_bloch_1998} can be proposed to rescale the anisotropy $K=K_u-K_d$ . This assumption neglects the stabilising long range character of the demagnetising energy. For a Néel spin structure, it also neglects the demagnetising contribution from the charged domain wall. This assumption will be discussed and compared to micromagnetic calculations (see section \ref{discussion_LPA}). 

	Treating demagnetising energy as a local anisotropy is correct for some recently proposed low-magnetisation systems such as ferrimagnetic films close to their compensation composition in the chosen working temperature range \cite{caretta_fast_2018} or synthetic  antiferromagnetic structures \cite{zhang_magnetic_2016}. 
	
 We assume a cylindrical symmetry for magnetisation and no z-dependence of the magnetisation because of the ultrathin (few atomic planes) thickness compared to the exchange length $l_{ex}=\sqrt{\frac{A}{K_d}}$. Magnetisation is $\vec M(r) = M_s \vec m$, where $\vec m$ is a unit vector and $M_s$ is uniform. Micromagnetic expressions are used to describe the film energy, so a continuous magnetisation field is assumed and we restrict the study to lengthscales larger than a few nm in order not to need an atomic model. However, our results agree with atomic spins on a lattice calculations at the small radius limit \cite{rohart_path_2016}. 
Due to the absence of a long range dipolar contribution, a relevant lengthscale for this problem is the Bloch length $l_B=\sqrt{\frac{A}{K}}$.

For large bubbles, $l_B$ scales with the Bloch domain wall width $\delta = \pi \sqrt{\frac{A}{K}}=\pi l_B$. The domain wall energy in the planar limit (large radius compared to width) for  a Bloch spin structure is $\gamma_{Bloch}=4\sqrt{AK}$, which becomes $\gamma_{DMI}=4\sqrt{AK} -\pi D$ when DMI is present assuming a Néel spin structure. It allows to define D$_c=\frac{4\sqrt{AK}}{\pi}$, which is the maximum value for D keeping a ferromagnetic ground state. In the absence of volume magnetic poles (local anisotropy assumption), no Neel to Bloch transition is expected when decreasing D. A Neel structure is favored as soon as D is non zero and its chirality is determined by the sign of D.

\subsection{Equations}
Following Kiselev et al.\cite{kiselev_chiral_2011}, micromagnetism energies are written using cylindrical coordinates and assuming a cylindrical symmetry.

The ferromagnetic exchange energy density is written using the usual isotropic term: $\epsilon_{ex}=A ((\frac{\partial \theta}{\partial r})^2+\frac{sin^2\theta}{r^2})$. Introducing the reduced radius $\rho=\frac{r}{l_{B}}$, the exchange energy is: 
$$E_{ex}= 2\pi A t\int_0^\infty((\frac{\partial \theta}{\partial \rho})^2+\frac{sin^2\theta}{\rho^2})  \rho d\rho$$
This exchange energy is scale-invariant, so E$_{ex}$ does not depend on the object radius. When it is the only energy, the minimiser is known, it is the Belavin-Polyakov (BP) skyrmion profile \cite{belavin_metastable_1975}. Its energy is $E_{ex}=8\pi A t$. Defining $r_0$ as the skyrmion radius when $m_z(r_0)$=0, the BP profile is $m_z=\frac{r_0^2-r^2}{r_0^2+r^2}$ and $m_r=\frac{2r_0r}{r_0^2+r^2}$. The $(\frac{\partial \theta}{\partial \rho})^2$ planar wall term and the $\frac{1}{\rho^2}$ curvature term contribute equally to the BP skyrmion exchange energy.\\
The magnetocrystalline anisotropy has an interfacial origin for ultra-thin films and it is modelled as a uniaxial one, with energy density $\epsilon_{mc}=K \sin^2\theta(r)$, which gives $$E_{mc}=2\pi A t  \int_0^\infty \sin^2\theta(\rho)\rho d\rho$$
In a cylindrically symmetric film, DMI is generated by the non equivalent interfaces between for example cobalt and the buffer (Pt) and the capping layer (AlO$_x$). The lack of inversion symmetry makes an assymetric exchange interaction possible. Using an atomic model, the energy can be written as $E_{DMI}=\vec D_{12} . \vec S_1 \wedge \vec S_2$, where $\vec D_{12}$ lies in the film plane, perpendicular to the $\vec r_{12}$ direction between both considered spins $\vec S_1$ and $\vec S_2$. It can be written as a micromagnetic energy \cite{bogdanov_thermodynamically_1994} : $$\epsilon_{DMI}=D (\frac{d\theta}{dr}+\frac{sin \theta cos \theta}{r})$$
The first term selects the wall chirality, the second term is due to the wall curvature and changes sign along the radial direction. It stabilises the skyrmion in the core part ($r<r_0$) but it is a positive energy contribution outside the core.\\
Normalising D to its maximum value allows to define $\xi=\frac{D}{D_c}=\frac{\pi D}{4\sqrt{AK}}$. We are interested in $\xi$ spanning the $[0, 1]$ range (i.e. 0\% to 100\%). We have :$$E_{DMI}=2\pi A t \int_0^\infty \frac{4\xi}{\pi}(\frac{d\theta}{d\rho}+\frac{sin \theta  cos \theta}{\rho}) \rho  d\rho $$

Finally the skyrmion total energy is :

\begin{equation} \label{Etot}
E=2 \pi A t \int_0^\infty[((\frac{\partial \theta}{\partial \rho})^2+\frac{sin^2\theta}{\rho^2}) +sin^2(\theta(\rho)) +\frac{4\xi}{\pi}(\frac{\partial \theta}{\partial \rho}+\frac{sin2\theta}{2\rho}) ]\rho d\rho
\end{equation}

E is calculated with respect to the perpendicularly saturated ferromagnetic state. Only demagnetising energy is non-zero and the energy density of the perpendicularly saturated film is $\epsilon=-K_d$. With our assumptions, ultrathin film (no z-dependence), cylindrical symmetry, no long range demagnetising term, only one equation must be solved and $\xi$ is the only parameter to determine the $\theta(\rho)$ profile. A numerical calculation was used to solve equation (\ref{Etot}) in the range $\xi=20\%$ to  $95\%$.\\

\begin{figure}
\begin{center}
\includegraphics[totalheight=8cm]{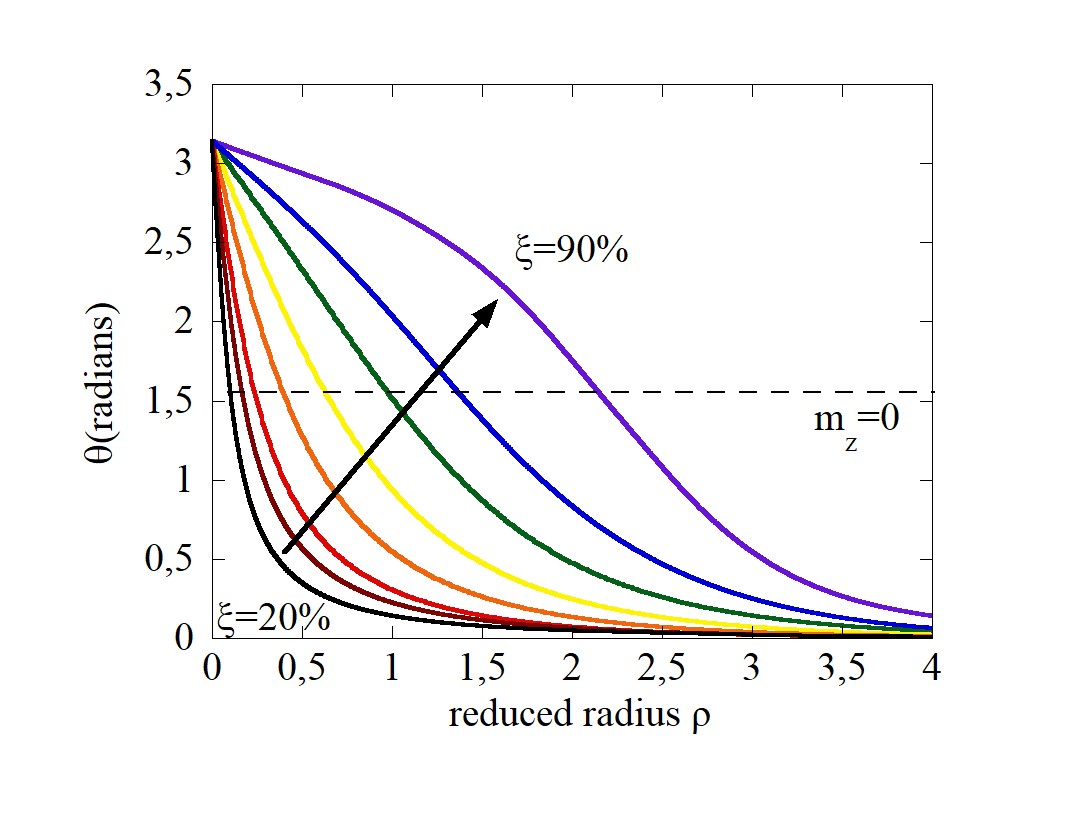}
\caption{\label{profiles} Magnetisation profile $\theta(\rho)$ of skyrmion solutions for $\xi$=20\% to 90\%, step 10\%.  }%
\end{center}
\end{figure}

\subsection {Numerical Solution}
The previous 1D equation was numerically solved using its Euler equation.

   Space is discretised over 10$^4$ nodes from $\rho=0$ to $\rho_{max}=$10 or 20. The boundary conditions are $\theta(0)=\pi$ and $\theta(\rho_{max})=2 e^{-\rho_{max}}$, which is the asymptotic tail for a 1D Bloch wall.  Energies were calculated in $2\pi A t$ units. The radius $\rho_0$ of the skyrmion is still defined when $m_z=0$, so $\theta(\rho_0)=\frac{\pi}{2}$. We define the skyrmion wall width $\delta$ from $\theta(\rho)$ as $\delta =- \frac{\pi l_B}{\frac{d\theta}{d\rho}(\rho_0)}$ 

\begin{figure}
\begin{center}
\includegraphics[totalheight=8cm]{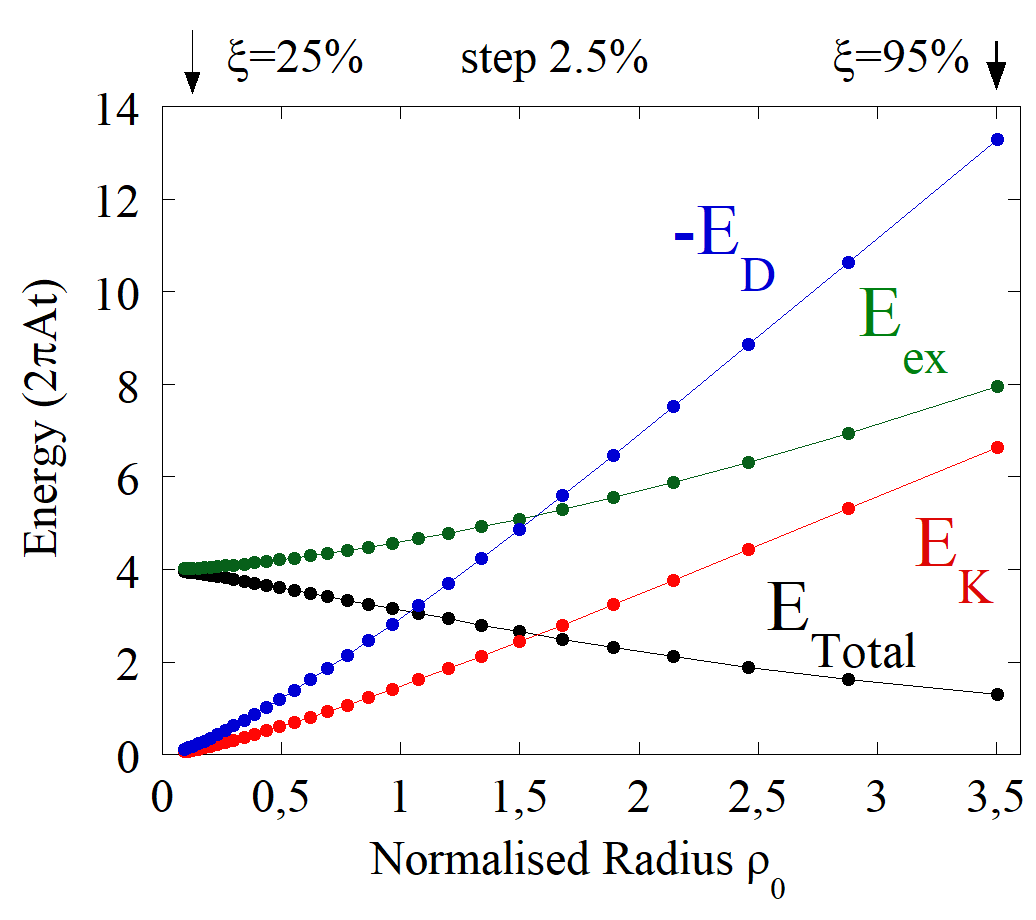}
\caption{\label{Energies} Energies (exchange, anisotropy, DMI and total) of the skyrmion solutions as a function of the skyrmion reduced radius. Equation (\ref{Etot}) was solved for the range $\xi$=[20\%-95\%], step 2.5\%.  }%
\end{center}
\end{figure}

\section{Results}

	The magnetisation profiles of some skyrmions are plotted in Fig(\ref{profiles}). As expected, a larger $\xi$ i.e., a larger DMI parameter D, stabilises a larger radius skyrmion. The magnetisation profiles evolves from a Belavin-Polyakov skyrmion profile, without a core, to a bubble-like skyrmion for larger $\xi$ with a core and a domain wall where $\theta(\rho)$ evidences an inflection point. A cycloidal structure becomes the ground state when D reaches $D_c$ ($\xi$=100\%). 

	The detailed energies of the equilibrium structures are plotted in Fig(\ref{Energies}) for $\xi$=[20\%-95\%].  The total energy decreases with increasing $\xi$ i.e., increasing D. Since D is a stabilising energy, increasing its value leads to a lower skyrmion energy even if the equilibrium size of the skyrmion is larger. In the calculated range, the equilibrium structure is always metastable compared to the ferromagnetic ground state where only the demagnetising energy contributes to the total energy. At large radius, it mirrors the fact that the effective domain wall energy stays positive whatever $\xi$ and at small radius, a finite (BP) exchange energy is required. Stable equilibrium structures with respect to the uniform perpendicularly magnetised state need long-range demagnetising stabilisation energy (multidomain demagnetised state).

\begin{figure}
\begin{center}
\includegraphics[totalheight=8cm]{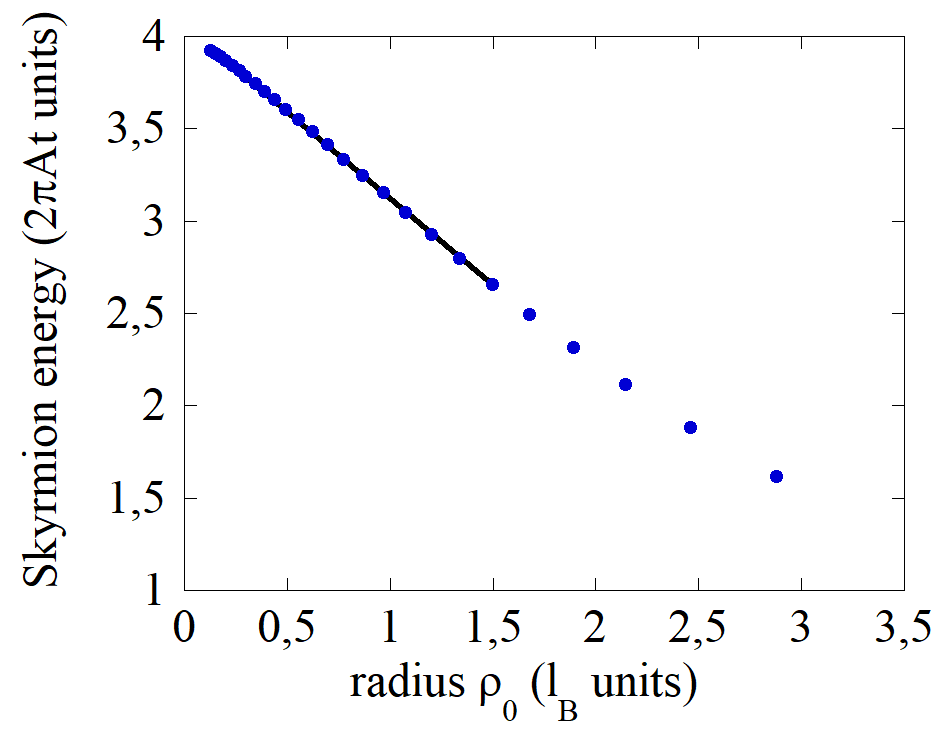}
\caption{\label{Energy} Energy of the skyrmion (in $2\pi At$ units) as a function of its radius $\rho_0$ (in l$_B$ units).  $\xi$ varies from =25$\%$ to 95$\%$. The continuous line is $E=4(1-0.231\rho_0)$}.%
\end{center}
\end{figure}

The exchange energy converges to $8\pi A t$ for small radii. This is the size-invariant BP skyrmion energy. Exchange and anisotropy energy tend to similar linear asymptotes for large radii. This asymptote is half the Bloch domain wall surface energy $\gamma_{Bloch}$, so it can be written as $2\sqrt{AK}2\pi rt=4\pi A t \rho_0$.

The DMI stabilising energy is quasi-linear with radius with a quadratic crossover for small radii. For a converged spin structure, the DMI energy to anisotropy energy ratio is -2, as theoretically expected (see proof, section \ref{discussion_DMIK}). The skyrmion total energy (Fig.\ref{Energy}) can be expressed in the range 20\% to 50\%D$_c$ as :
$$E = 8 \pi A t ( 1 - 0.231 \rho_0)$$

\begin{figure}
\begin{center}
\includegraphics[totalheight=8cm]{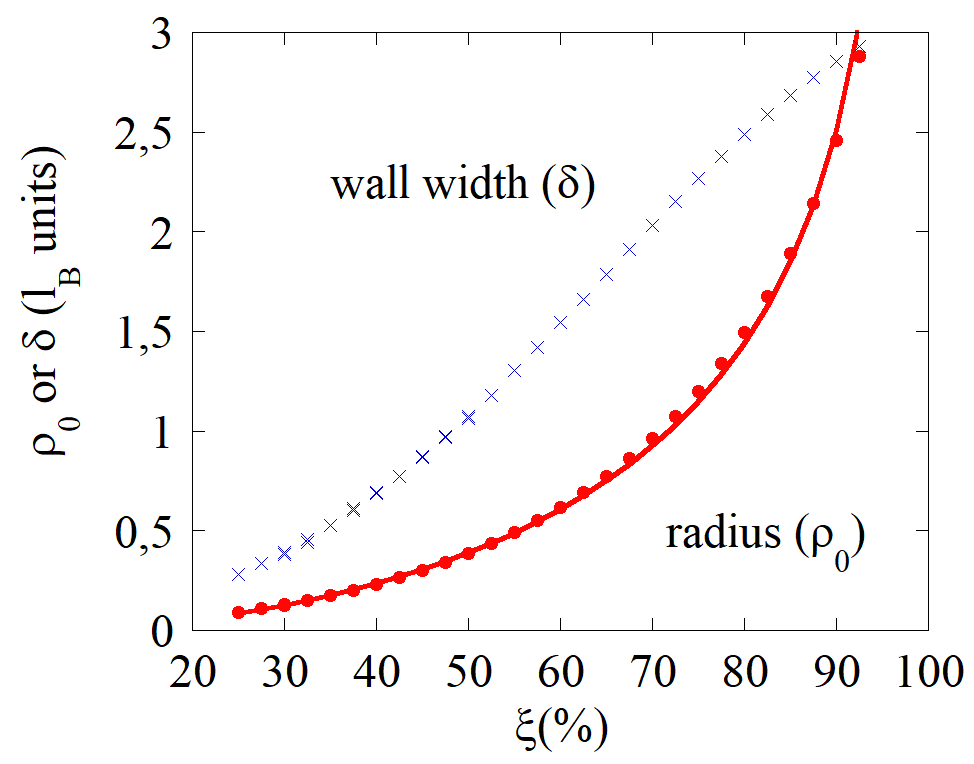}
\caption{\label{radius}Skyrmion reduced radius $\rho_0$ (full circle) and skyrmion wall width $\delta$ (cross) as a function of $\xi$. The continuous line is $\rho_0=1.35\xi^2/\sqrt(1-\xi^2)$.}%
\end{center}
\end{figure}		

Fig. \ref{radius}	shows the reduced skyrmion radius $\rho_0$ and the wall width $\delta$ as a function of $\xi$. The wall width $\delta$ evolves from $\delta=2\rho_0$ for a BP profile (small $\xi$) up to the Bloch domain wall width $\delta=\pi$ (in $l_B$ unit) for larger objects.\\

The skyrmion reduced radius can be expressed as $\rho_0=1.35\frac{\xi^2}{\sqrt{1-\xi^2}}$, which is correct within 5\% in the range $\xi=20\%$ to $90\%$.

\section{Discussion}
\subsection{Theoretical DMI to anisotropy energy ratio}	\label{discussion_DMIK}
As mentioned previously a simple relation between calculated DMI and anisotropy energies is found. Here is the proof.	Let's assume that $\theta(\rho)$ minimises the total energy (Equation \ref{Etot}), which can be written :
$$E = E_{ex} + E_K + E_D$$ 
Let's homothetically expand radially the spin structure by a factor $\lambda$ : $\theta(\rho')=\theta(\lambda \rho)$. The total energy becomes $E' = E_{ex} + \lambda^2 E_K + \lambda E_D$. It must be larger than the original energy which was that of the optimised spin structure. $E'(\lambda)$ is minimum for $\lambda = \frac{-E_D}{2 E_K}$ and it should correspond to our original spin structure, so $\lambda=1$.

			Finally one has that $E_D = -2 E_K$ i.e. the integrated DMI energy is always twice the opposite to the anisotropy energy. Since the proof is based on homogeneity, adding higher order anisotropy terms will give the same relation. This identity can be used to assess numerical models. Note that this result is correct for the integrated energies but not for the local energy densities. Bogdanov et al.\cite{bogdanov_properties_1994} has a similar result for a thick magnetic vortex, invariant along z, which is mathematically equivalent.\\

\subsection{Local Anisotropy Approximation} \label{discussion_LPA}
\begin{figure}
\begin{center}
\includegraphics[totalheight=8cm]{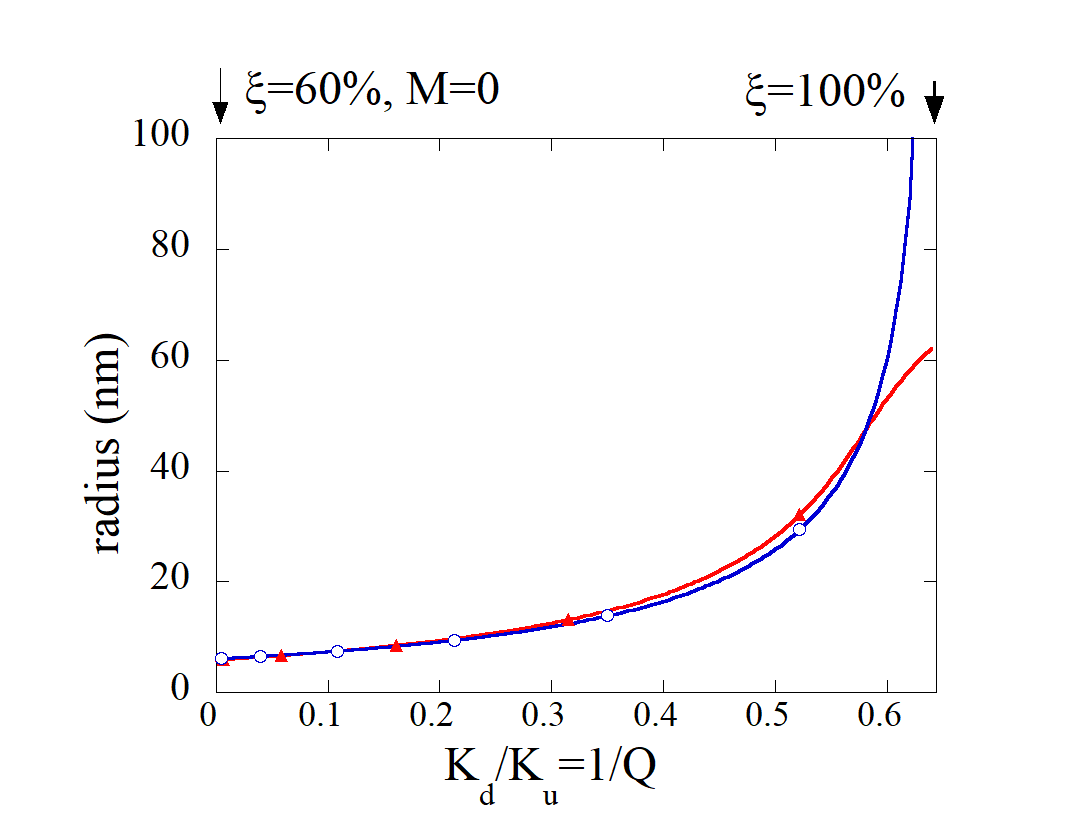}
\caption{\label{MuMax1} Skyrmion radius calculated as a function of the film spontaneous magnetisation M$_s$ keeping all parameters constant. Mumax3 micromagnetic solver (red triangles) and our 1D model (blue circles) with effective local anisotropy $K=K_u-K_d$. The x-axis is $M_s^2$ normalised to be 1/Q. The chosen material parameters are A=10 pJ/m, K$_u$=10$^5$ J/m$^3$, D=0.76mJ/m$^2$, i.e. $0.6D_c$ when $M_s$=0 and t=1nm. Increasing $M_s$ corresponds to decreasing K and $D_c$ and increasing $\xi$ (top scale).  }
\end{center}
\end{figure}

In order to take into account demagnetising terms, without calculating them, the local anisotropy approximation was used to rescale anisotropy. In order to test the validity of this approach for large magnetisations, a micromagnetic solver (Mumax3) \cite{vansteenkiste_design_2014, mulkers_effects_2017} has been used. Fig. \ref{MuMax1} represents the equilibrium skyrmion radius for t=1nm, A=10pJ/m, K$_u$=10$^5$J/m$^3$, D=0.76mJ/cm$^2$, calculated on a 512x512 mesh with 1nm isotropic mesh cellsize. The blue line is the 1-D model with an effective anisotropy $K=K_u-K_d$. When Q is larger than 3, it superimposes the full Mumax calculation. When $M_s$ increases, the effective anisotropy K decreases which makes $D_c$ decrease. Since D is constant, $\xi$ goes to 100\% and the skyrmion radius diverges before reaching the in-plane to out-of-plane magnetic transition (Q=1). Boundary effects are present in the MuMax3 calculation (512nmx512nm simulation box). Part of the finite size effect is corrected by adding as an external field the demagnetising field due to the saturated magnetisation out of the simulated box. However, repulsion between the skyrmion and the DMI tilted spins at the box edges are still present and limit the radius increase. Up to demagnetising anisotropies $K_d$ which are significant compared to $K_u$, the analytical model is a useful quantitative model expanding its use to non-zero-magnetisation systems.

\subsection{Neglecting volume charges} \label{discussion_Bloch}
Representing the demagnetising energy as a local anisotropy term means neglecting the volume charges that are associated to the Neel-domain wall spin structure. Since the effective K model agrees with the full demagnetising calculation (micromagnetic solver), it shows that neglecting the volume charges does not modify significantly the spin structure. The micromagnetic solver allows to extract in plane $H_{dr}$ and out-of-plane $H_{dz}$ components of the demagnetising field. By extracting the $H_{dz}$ and $H_{dr}$ components for a 10 nm skyrmion, as expected from the radius/thickness ratio, there is one order of magnitude between them since $r_0 >> 10t$ and $H_{dr}$ is a minor correction. [not shown].

When the magnetisation is finite, our model neglects the expected Néel to Bloch transition when D decreases. It happens when $E_{DMI}$ and the volume charge contribution to $E_{demag}$ compensate.
An estimate for $E_{DMI}$ is :  $E_{DMI}= D\frac{\pi}{2r_0}V$, where $D\frac{\pi}{2r_0}$ is an estimate for the DMI density of energy and $V$ is the skyrmion volume. An estimate for $E_{demag}$(Néel) is $K_d \frac{t}{r_0}V$, where $\frac{t}{r_0}$ is a demagnetising factor for the in-plane demagnetising field.
A cross over is expected for $D_{N-B}=\frac{2}{\pi}K_dt \approx 0.6 K_d t$.
Recently a more quantitative prediction $D_{N-B}=0.92 K_d t$ using a more elaborate calculation\cite{bernand-mantel_unraveling_2020} has been reported. It is worth noticing that our rough estimate identify the relevant physics and provides the numerical prefactor within a factor 2. So, when M$_s$ increases, or when D decreases, a Neel to Bloch spin structure transition is expected. Unlike the planar domain wall case\cite{thiaville_dynamics_2012}, where the wall keeps its length across the Bloch to Néel transition, such a transition for a skyrmion could be associated with a collapse of the spin structure if the stabilising demagnetisaton energy does not compensate the spin structure energy cost. For small values of D, this Bloch-Néel spin structure transition is expected and our analytical model should not be used anymore.

\subsection{Design Rules}
We will now apply the model to define design rules for stable room temperature 10-nm-diameter skyrmions. Since $r_0=\rho_0 l_B=5$nm, and $\rho_0$ is typically of order 1, a small Bloch length of 5-20 nm is desirable. The study of the nucleation and collapse energies will help to adjust the choice of material parameters.

First, the nucleation energy allows to create the skyrmion. Starting from a ferromagnetic uniform state, the first skyrmion nucleus requires a finite exchange energy $8\pi A.t$ even for a vanishingly small nucleus. This energy does not scale with $r_0$ for small radii. In the atomic limit, which should be considered to describe the initial nucleus, this conclusion still holds\cite{rohart_path_2016}. Taking typical values A=10 pJ/m and t=1nm gives $8\pi A. t = 60 kT_{300K}$. Such a nucleation barrier is large enough to prevent spontaneous nucleation from thermal fluctuations at room temperature. Assisted nucleation using a heat source (current pulse and associated Joule effect) or a magnetic field pulse (Zeeman pressure) may be required. The nucleation barrier could also be decreased using for example an electric gate \cite{srivastava_large-voltage_2018}. A nucleation pad with reduced thickness (lower t but also lower A due to finite size effect for nanometer-thick films) could also be used to create the skyrmion which could then be displaced to the relevant region using a spin-polarised current induced motion.   

The collapse barrier is the difference between the skyrmion metastable energy and the energy maximum at small radii.
$$E_{collapse}=8\pi A. t - 8\pi A. t (1- 0.23\rho_0)=8\pi A. t.0.23\rho_0$$
Choosing a lifetime $\tau$ of 1 second for operational use of the skyrmion and using an Arrhenius law:
$$ \tau = \tau_0 e^{\frac{\Delta E}{kT}}$$
The try time $\tau_0$ is usually taken in the 0.1-1 ns range\cite{rohart_path_2016}, which gives a barrier $\Delta E = 20 kT$\\
Which brings  $0.23 \rho_0$ to be around 1/3, i.e. $\rho_0 \approx 1$. For 10-nm diameter skyrmions, i.e. $r_0=5$nm, it implies $l_B=5$ nm.
Using typical material properties, A=10 pJ/m and K=4.10$^5$J/m$^3$ give $l_B=\sqrt{\frac{A}{K}}=5$ nm. So materials with large perpendicular anisotropies are preferred. As regards DMI, small $\xi$ are preferred to get small structures. Given the previous choice of exchange and anisotropy constants, D should be kept smaller than D$_c$ which is 2.5 mJ/m$^2$ with these A and K values.  

What are the limits to these conclusions ? For 10-nm-diameter skyrmions, the local character of the demagnetising field is still a good assumption. The assumption of circular shape is even more valid for smaller objects where fluctuations would be at the nm scale which represents large exchange costs. Comparison to calculations performed on an atomic lattice show that the micromagnetic conclusions do not collapse when sizes are typically 5 atomic planes in thickness and 50 spins in diameter (typically 10nm). A more difficult assumption is the uniform character of all magnetic properties. Property fluctuations are less averaged out on a 10nm scale. Due to the polycrystalline nature of most metallic layers, 10-nm skyrmions would experience material fluctuations at a similar scale (grain size) and grain boundaries could become efficient pinning sites.

 Zeeman energy (applied magnetic field) was not considered in this work. When $M_s$ is large enough and depending on the choice of a stabilising or destabilising applied magnetic field direction, Zeeman energy could be added to manipulate the skyrmion. It is a good external parameter to modify the radius and to control the density of the objects.

\section{Conclusion}
	The model we developed allows to discuss the size, profile, nucleation and collapse barriers of DMI-stabilised skyrmions not using a 5-degrees of freedom phase diagram (A, K, D, M$_s$, t) but simply 2 parameters, the Bloch length and the normalised DMI. This gives a simpler picture of the role of the material parameters, with physical arguments to discuss all trends. As a consequence, simple rules to design skyrmions with certain size and stability are proposed. The model agrees quantitatively with existing models and experimental results.

\section*{Acknowledgements}
\paragraph{Funding information}
This project has received funding from the DARPA TEE program. M.A.M. acknowledges an Erasmus scholarship from Universidad Autonoma de Madrid to Université Grenoble Alpes.


\bibliography{Biblio_M0_c}

\nolinenumbers

\end{document}